\documentclass[letterpaper, 
              ]{jacow}
\makeatletter%
	\ifboolexpr{bool{xetex}}
	 {\renewcommand{\Gin@extensions}{.pdf,%
	                    .png,.jpg,.bmp,.pict,.tif,.psd,.mac,.sga,.tga,.gif,%
	                    .eps,.ps,%
	                    }}{}
\makeatother

\ifboolexpr{bool{xetex} or bool{luatex}} % test for XeTeX/LuaTeX
 {}                                      % input encoding is utf8 by default
 {\usepackage[utf8]{inputenc}}           % switch to utf8

\usepackage[UKenglish]{babel}			 
\usepackage[final]{pdfpages}
\usepackage{multirow}
\usepackage{ragged2e}
\usepackage{subfigure}
\begin{document}

\title{RECENT RESULTS FROM MICE ON MULTIPLE COULOMB SCATTERING AND ENERGY LOSS}

\author{P. Franchini\thanks{p.franchini@warwick.ac.uk},~University of Warwick,~Coventry,~UK\\
        on behalf of the MICE Collaboration}
        
\maketitle

\begin{abstract}
  Muon beams of low emittance provide the basis for the intense, well 
  characterised neutrino beams of a neutrino factory and for multi-TeV lepton-antilepton 
  collisions at a muon collider. The international Muon Ionization Cooling Experiment (MICE) will demonstrate 
  ionization cooling, the technique by which it is proposed to reduce the 
  phase-space volume occupied by the muon beam. MICE was constructed in a 
  series of steps. Data were taken in 2016 and 2017 in the Step~IV configuration which was optimised for studying the properties of liquid hydrogen~(LH$_2$) 
  and lithium hydride~(LiH). Preliminary results from ongoing analyses will be described.
\end{abstract}

%%%%%%%%%%%%%%%%%%%%%%%%%%%%%%%%%%%%%%%%%%%%%%%%%%%%%%%%%%%%%%%%%%%%%%%%%%%%%%%%%%%%%%%%%%

\section{Introduction}
%Accelerators and ionization cooling
Muon colliders and neutrino factories will require stored muons with high intensity and low emittance~\cite{geer}.
Muons are produced as tertiary particles ($p+N \to \pi+X$, $\pi \to \mu+\nu$) inheriting a large emittance (volume of the beam in the position and momentum phase space). For efficient acceleration, the phase-space volume of these beams must be reduced significantly (``cooled''), in order to be accepted by traditional accelerator components.
Due to the short muon lifetime, ionization cooling is the only practical and efficient technique to cool muon beams~\cite{neuffer}.
In ionization cooling, the muon beam loses momentum in all dimensions by ionization energy loss when passing through an absorbing material, reducing the RMS emittance ($\varepsilon_\mathrm{RMS}$) and increasing its phase space density. Subsequent acceleration though radio frequency cavities restores longitudinal energy, resulting in a beam with reduced transverse emittance.
A factor of 10$^5$ in reduced 6D emittance has been achieved in simulation with a 970~m long channel~\cite{stratakis}.

The rate of change of the normalized transverse RMS emittance $\varepsilon_\mathrm{N}$ is given by the ionization cooling equation~\cite{fernow}:
\begin{equation}
  \frac{d\varepsilon_\mathrm{N}}{ds}\simeq  
  -\frac{\varepsilon_\mathrm{N}}{\beta^{2}E_{\mu}}\left\langle\frac{dE}{ds}\right\rangle + 
  \frac{\beta_\mathrm{t}(13.6~[\mathrm{MeV}])^2}{2\beta^3E_{\mu}m_{\mu}X_0}
\label{eq:emittance}
\end{equation}
where $\beta c$ is the muon velocity, $\langle dE/ds \rangle $ is the average rate of energy loss, $E_\mu$ and $m_\mu$ are the muon energy and mass, $\beta_\mathrm{t}$ is the transverse betatron function and $X_0$ is the radiation length of the absorber material.
The first term on the right can be referred as the ``cooling'' term given by the ``Bethe equation'',
while the second term is the ``heating term'' that uses the PDG approximation for the multiple Coulomb scattering.

\begin{figure*}
  \begin{center}
    \includegraphics[width=0.8\textwidth]{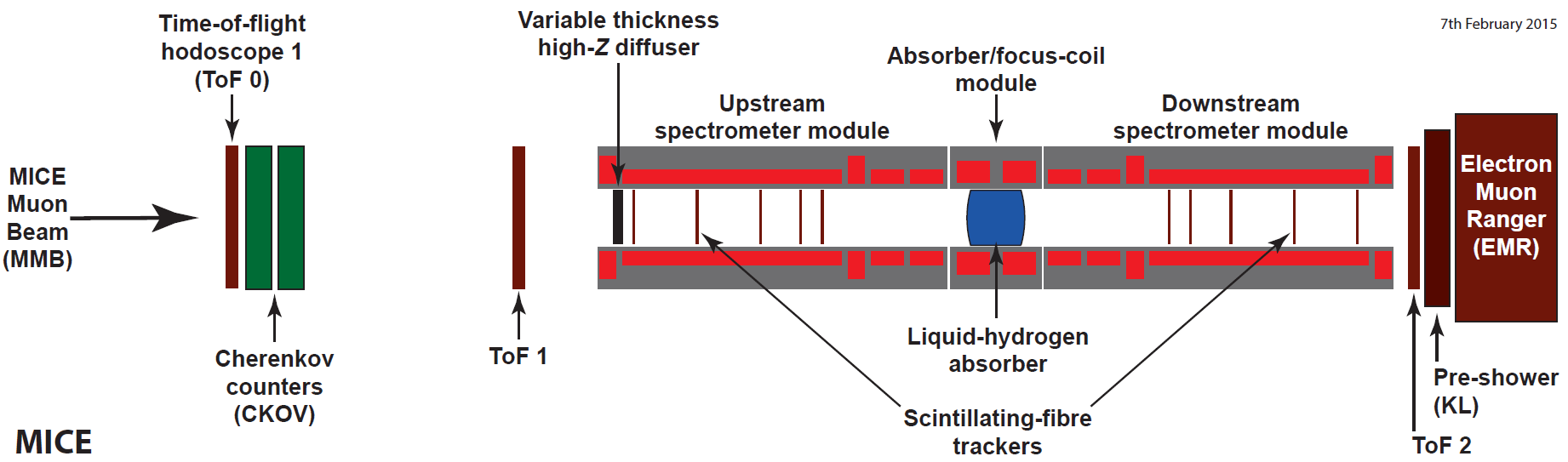}
    \caption{Layout of MICE Step IV configuration, showing the absorber, tracking spectrometers and detectors for particle identification.}
    \label{fig:beamline}
  \end{center}
\end{figure*}

%MICE Step IV
A schematic drawing of MICE Step IV is shown in Fig.~\ref{fig:beamline}.

%Beam line and detectors
MICE is instrumented with a range of detectors used for particle identification and position-momentum measurement.
This includes a scintillating fibre tracker upstream and downstream of the absorber placed in a strong solenoid field to measure the position and the momentum (with a spatial resolution around 0.3~mm). MICE is also equipped with a series of particle identification detectors, including 3 time-of-flight hodoscopes (ToF0/1/2, with a time resolution around 60~ps), 2 threshold Cherenkov counters, a pre-shower calorimeter and a fully active scintillator calorimeter.
% citation?

%Step IV measurements
MICE data taking was concluded in December 2017 (in the Step IV configuration) in order to make detailed measurements of
multiple Coulomb scattering and energy loss of muon beams at different momenta and channel configurations, with lithium hydride and liquid hydrogen absorbers.
The collaboration also seeks to measure the reduction in normalized transverse emittance~\cite{blackmore}, comparing the emittance of a sample of muons selected in the upstream tracker with the emittance of the same sample measured in the downstream one, after passing through the absorber.

%%%%%%%%%%%%%%%%%%%%%%%%%%%%%%%%%%%%%%%%%%%%%%%%%%%%%%%%%%%%%%%%%%%%%%%%%%%%%%%%%%%%%%%%%%

\section{Measurements of scattering distributions}
% check JN note
% update plots

Though multiple Coulomb scattering is a well understood phenomenon, results from \mbox{MuScat~\cite{muscat1}\cite{muscat2}} indicate that the effect in low $Z$ materials is not well modelled in simulations such as GEANT4~\cite{geant4}.
MICE will therefore measure the multiple Coulomb scattering distribution to validate the scattering model and understand the heating term in Eq.~\ref{eq:emittance}, in order to make more realistic predictions of the emittance reduction.
Both data with field off and field on in the scintillating fibre tracker are available for MICE. While the field on data is still being analysed, the field off analysis is presented here.

\begin{figure}[ht!]
  \begin{center}
  \begin{subfigure}
  \centering
  \includegraphics[width=0.7\columnwidth]{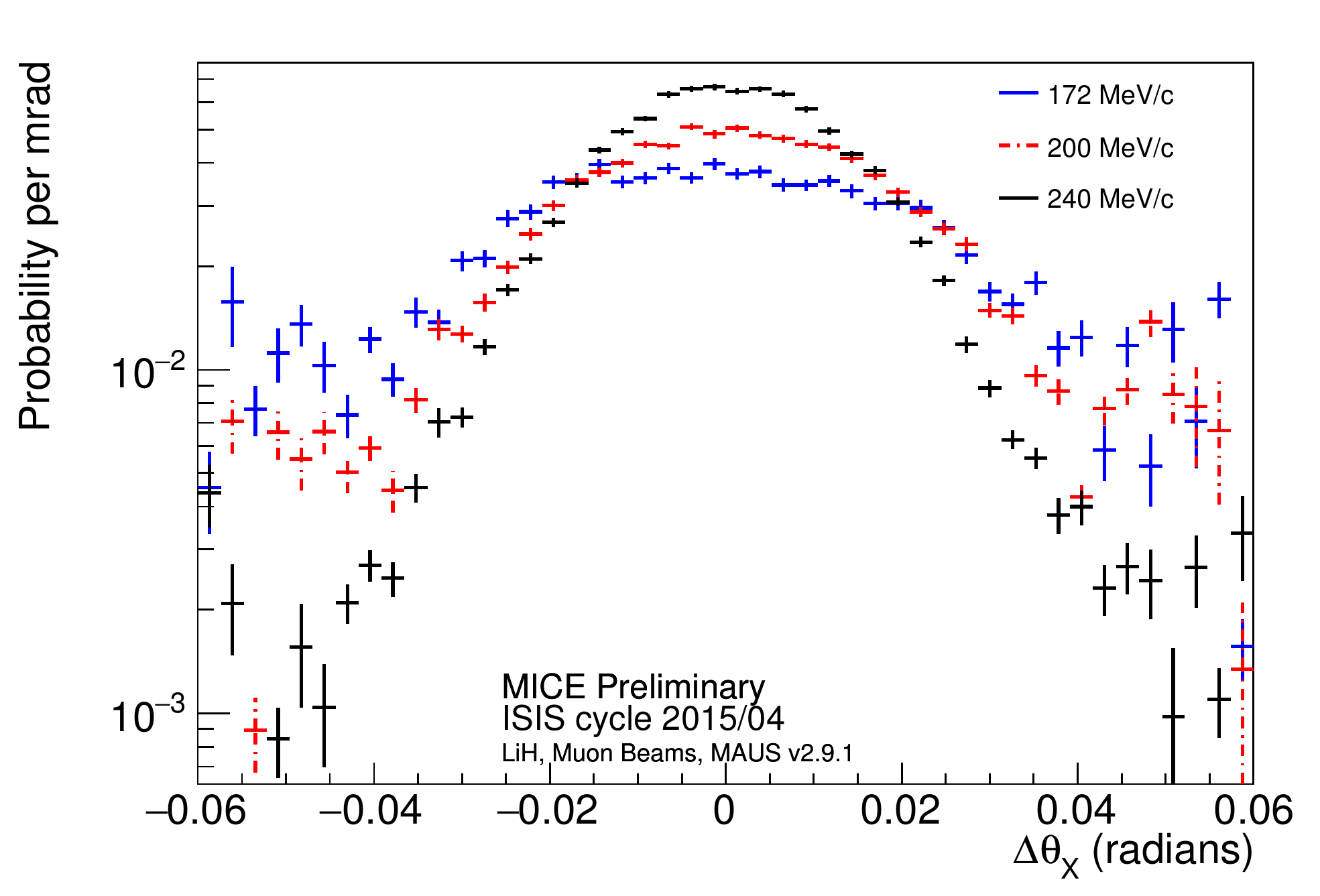}
  \end{subfigure}
  \begin{subfigure}
  \centering
  \includegraphics[width=0.7\columnwidth]{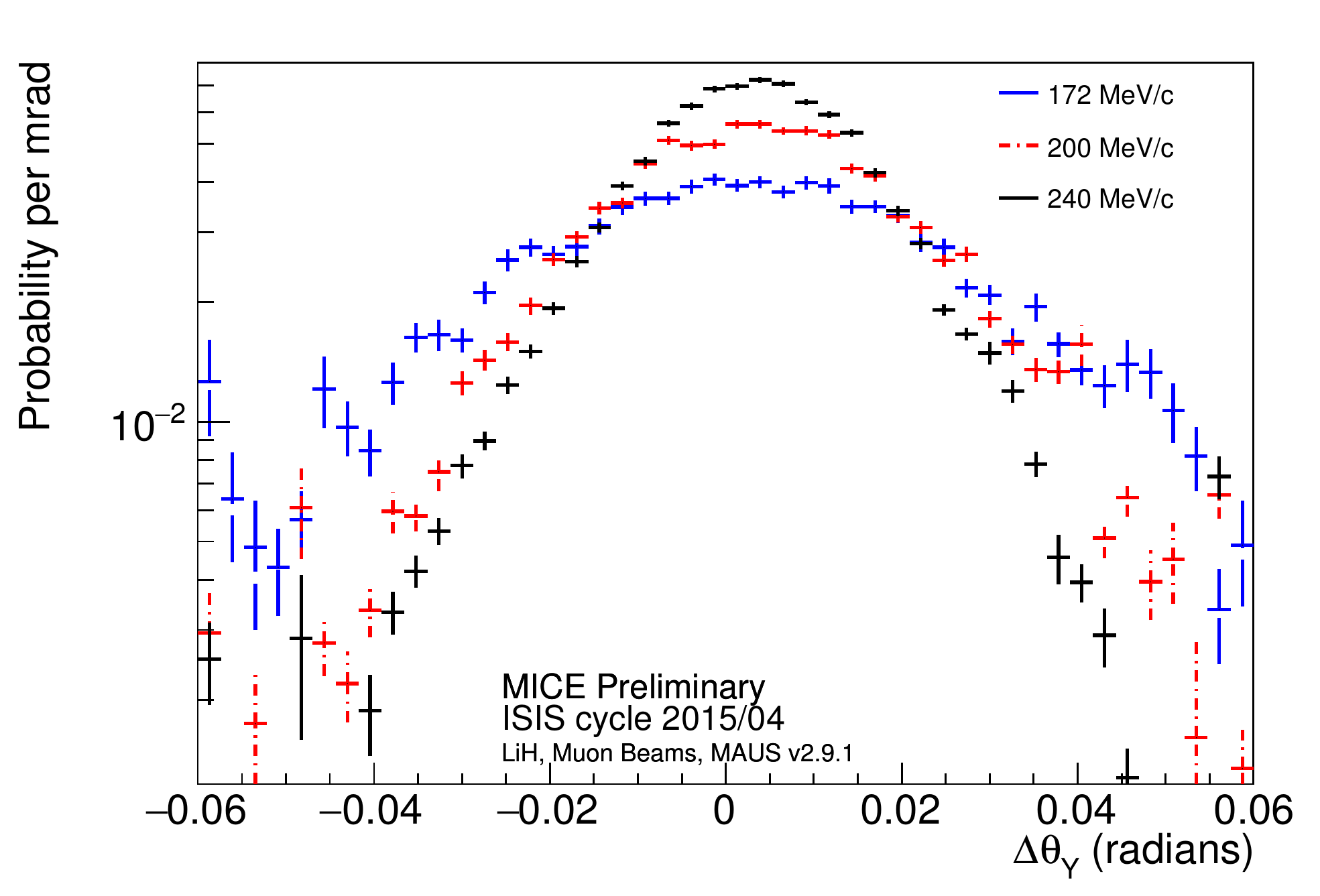}
  \end{subfigure}
  \caption{Scattering distributions projected in the Y-Z (top) and X-Z (bottom) planes of muons passing through the LiH absorber.}
  \label{fig:scattering}
  \end{center}
\end{figure}

MICE has collected data for muon beams at three different momenta, 172~MeV/c (in order to compare with \mbox{MuScat}), 200~MeV/c and 240~MeV/c
with and without the LiH absorber disk in place (thickness 65~mm, $X_0=79.62~\mathrm{g~cm^{-2}}$) and with a full and an empty liquid hydrogen absorber vessel ($\sim$~22~litre). Here the LiH analysis will be presented.

The position and momentum of each muon is measured by the trackers along with the time-of-flight, the latter also provides particle identification.
Selection criteria were imposed on each track to select a well understood  sample.
Bayesian deconvolution was applied to the selected data in order to extract the scattering distribution within the absorber material and comparisons have been made to GEANT4, to the Moliere model~\cite{moliere} as well as to a stand alone scattering model developed by Carlisle and Cobb~\cite{cc}.
Data taken with LiH on a full range of beams, deconvolved using the GEANT model, are shown in Fig.~\ref{fig:scattering}.
% errors:
Different contributions to the systematic uncertainty have been considered: sensitivity to the thickness of the absorber, time of flight cuts used for momentum selection, alignment of the detectors and choice of the fiducial cuts. The time of flight systematics dominate.
% results:
%% width for the central momenta 
The scattering width taken from the scattering distributions projected in the X-Z and Y-Z planes are
$\Theta$~=~23.3~$\pm$~0.9~$\pm$~0.2~mrad at 172~MeV/c,
$\Theta$~=~17.9~$\pm$~0.4~$\pm$~0.5~mrad at 200~MeV/c and
$\Theta$~=~14.2~$\pm$~0.1~$\pm$~0.5~mrad at 240~MeV/c in LiH.
The preliminary analysis indicates that GEANT4 underestimates the scattering width, while the PDG model overestimates it~\cite{nufact}.

\begin{figure}[ht!]
  \begin{center}
  \begin{subfigure}
  \centering
  \includegraphics[width=0.7\columnwidth]{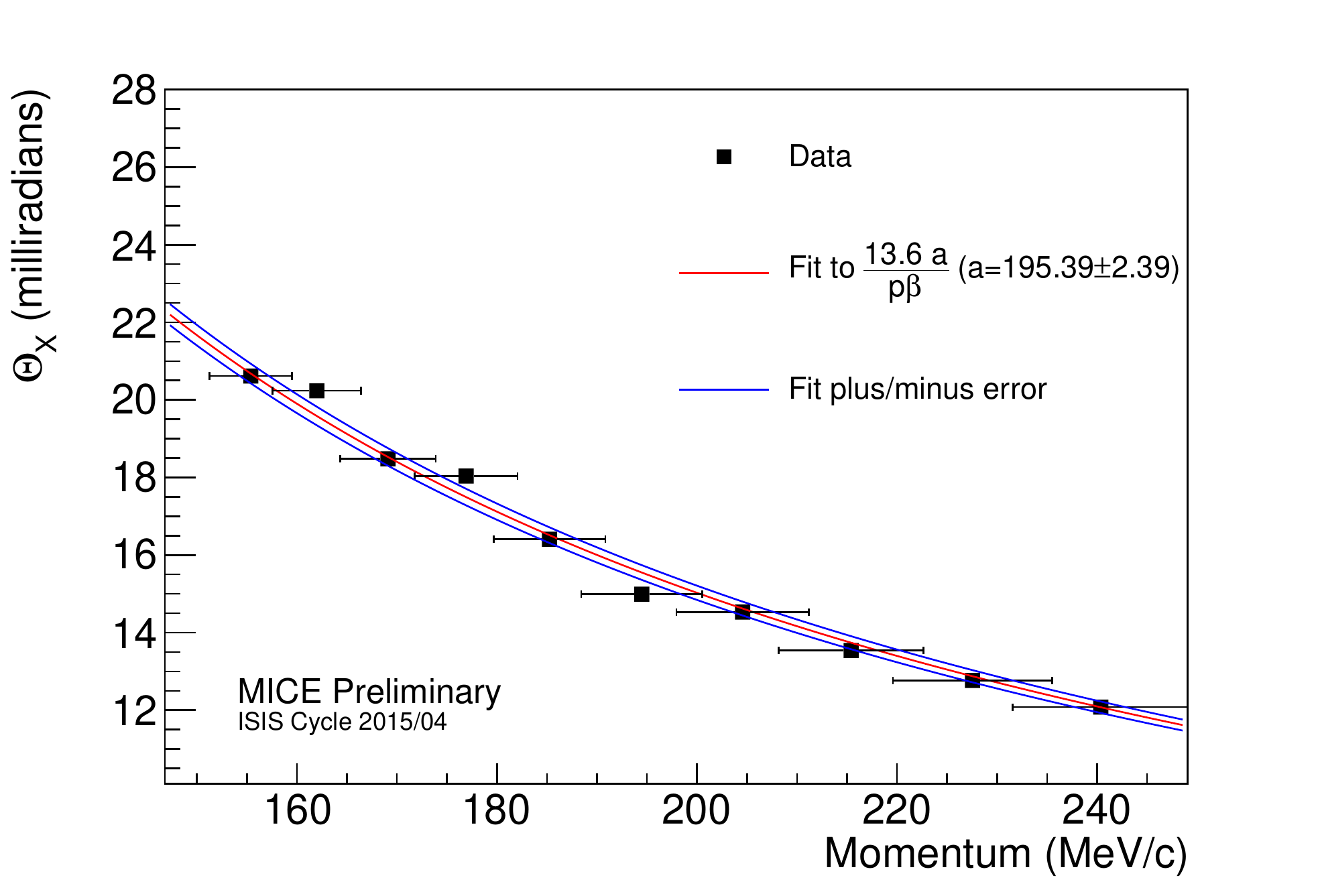}
  \end{subfigure}
  \begin{subfigure}
  \centering
  \includegraphics[width=0.7\columnwidth]{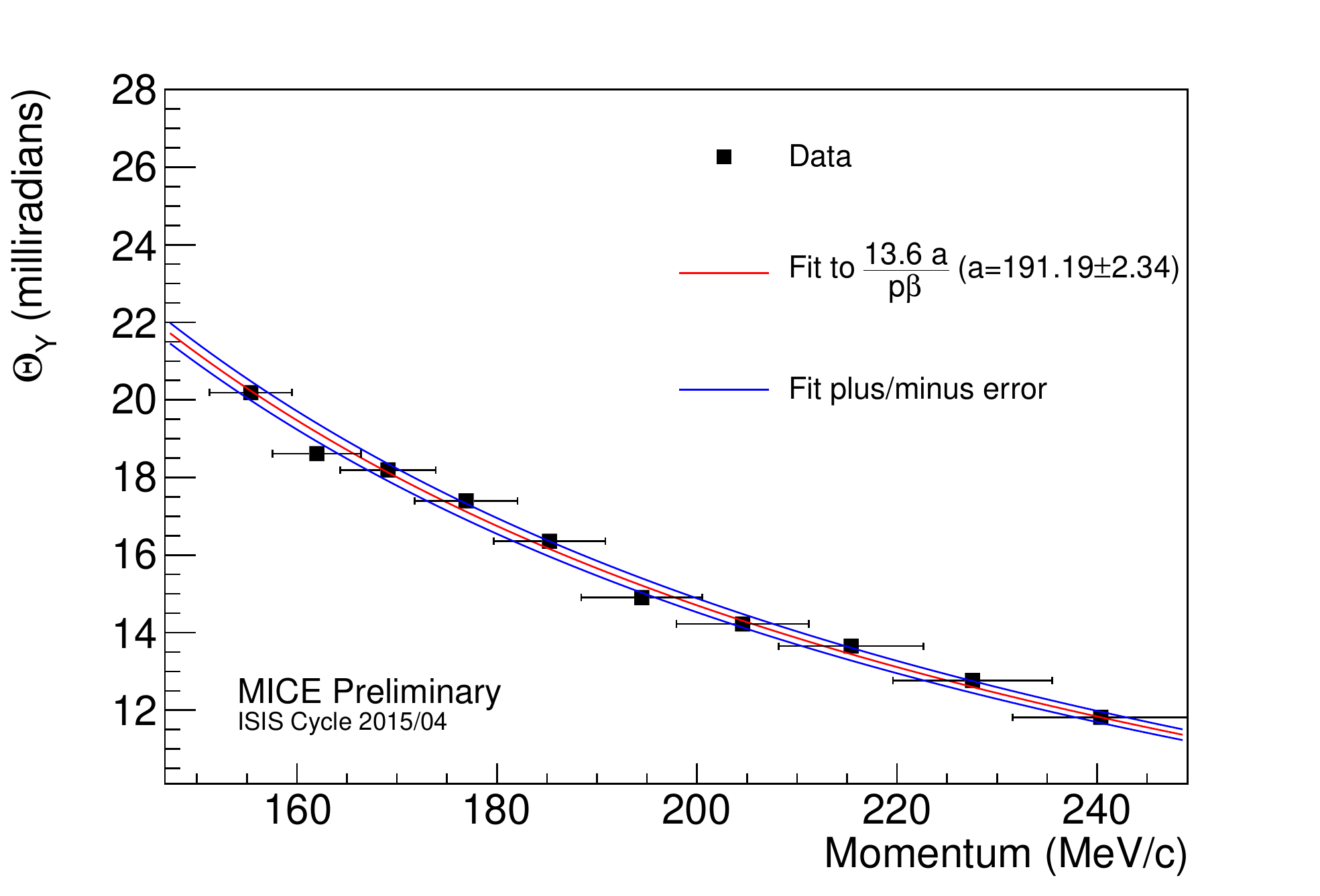}
  \end{subfigure}
  \caption{Scattering analysis repeated at each momentum bin, projected in the Y-Z (top) and X-Z (bottom) planes. The fit to the data is shown.}
  \label{fig:scattering_dependent}
  \end{center}
\end{figure}

%% momentum dependence of scattering:
Data were collected over a wide range of momenta. The data were then binned in momentum and the analysis repeated for each bin with the scattering width determined in each case.
The distribution has been fitted with an expression with a $1/p\beta$ dependence of a similar form to the Rossi and Greisen expression for the RMS scattering~\cite{rossi} recommended by the PDG~(Fig.~\ref{fig:scattering_dependent}).
%The projected scattering angles are consistent with the PDG predictions as shown in Fig.~\ref{fig:scattering_dependent}.

%%%%%%%%%%%%%%%%%%%%%%%%%%%%%%%%%%%%%%%%%%%%%%%%%%%%%%%%%%%%%%%%%%%%%%%%%%%%%%%%%%%%%%%%%%

\section{Measurement of energy loss}
% check workshop talk
% update/add plots
% add more text

The mean rate of energy loss for relativistic charged heavy particles traversing matter is given by the ``Bethe equation''~\cite{bethe}:
\begin{equation}
- \left\langle \frac{dE}{dX} \right\rangle =
 K z^2 \frac{Z}{A} \frac{1}{\beta^2} \left[ \frac{1}{2} \ln{\frac{2 m_\mathrm{e} c^2 \beta^2 \gamma^2 W_{\mathrm{max}}}{I^2}} - \beta^2 - \frac{\delta(\beta\gamma)}{2}\right]
\label{eq:bethe}
\end{equation}
where the mean excitation energy, $I$, in hydrogen is known at the 5\% level but has never been measured in lithium hydride. Small differences are expected between the energy loss in LH$_2$ and in LiH.

\begin{figure}
  \begin{center}
    \includegraphics[width=.8\columnwidth]{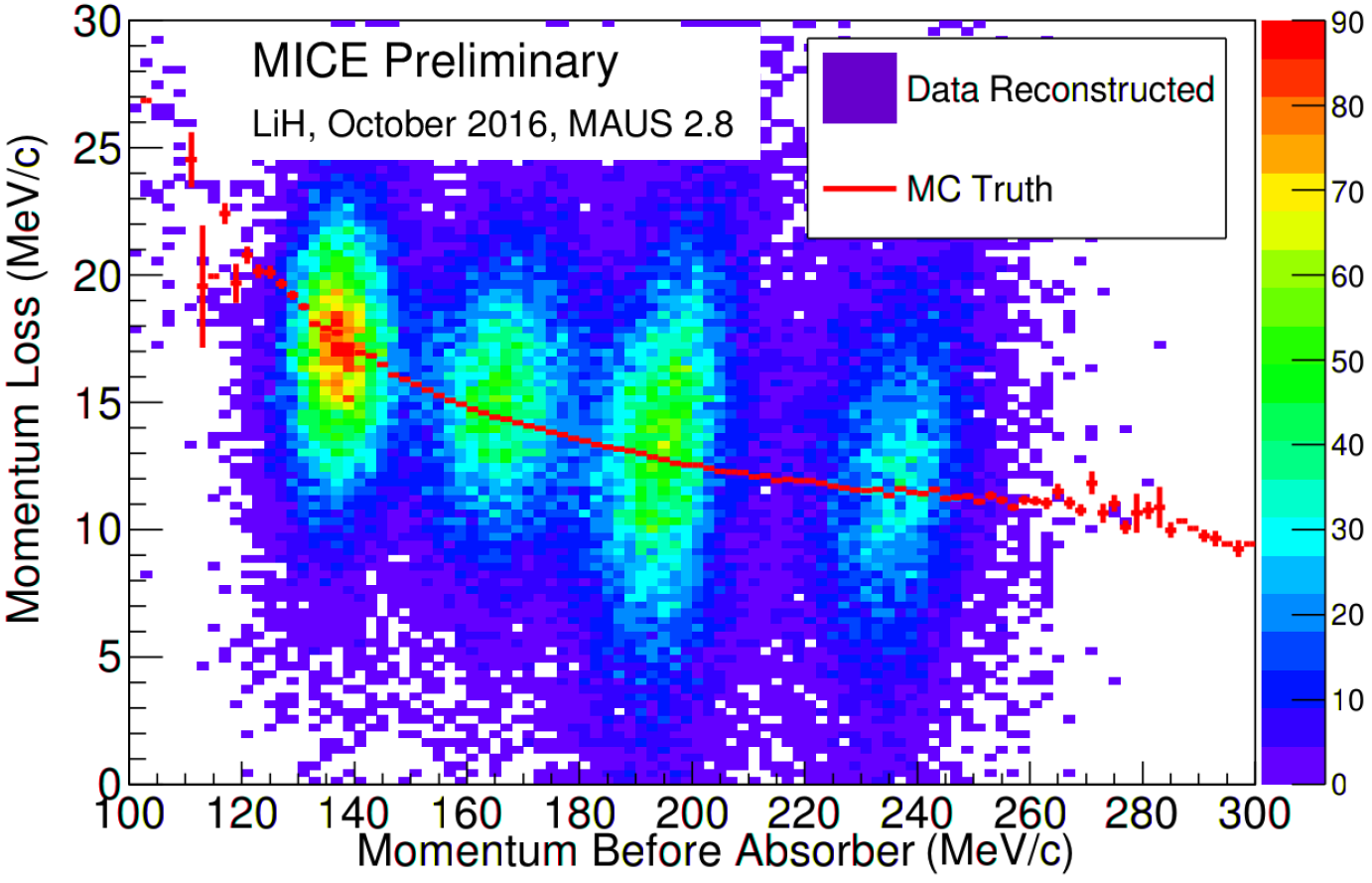}
    \caption{Distribution of the reconstructed momentum loss across the absorber for several different beams compared with the MC truth.}
    \label{fig:energy_loss}
  \end{center}
\end{figure}

MICE measures the momentum upstream and downstream of the absorber using information from the trackers combined with measurements of the time of flight.
Data has been analysed using central muon-beam momenta of 140, 170, 200 and 240~MeV/c in the presence of 3~T magnetic fields, with and without the LiH absorber~(Fig.~\ref{fig:energy_loss}).
Preliminary results for 200 MeV/c muons in magnetic field traversing the LiH absorber show that the mean momentum loss is \mbox{$\Delta p = 12.8 \pm 5.3$~MeV/c}.
%agrees with GEANT4 simulation ($\Delta p = 12.8 \pm 5.3$~MeV/c).
Further studies are planned to deconvolve the energy loss measured without absorber from the measurement with the absorber in order to obtain the energy loss in the absorber.
The final goal will be to measure the correlation between energy loss and multiple Coulomb scattering.

%%%%%%%%%%%%%%%%%%%%%%%%%%%%%%%%%%%%%%%%%%%%%%%%%%%%%%%%%%%%%%%%%%%%%%%%%%%%%%%%%%%%%%%%%%

\section{Conclusions}
MICE, at Step IV, has successfully collected all the required data in order to measure the properties of liquid hydrogen and lithium hydride that affect the performance of an ionization cooling channel. Step~IV data taking commenced in 2016 and concluded at the end of 2017: scattering and energy loss measurements on LiH and LH$_2$ have been performed.
The Step IV configuration will also be used to study the effect of channel optics and of the input beam momentum and emittance on the ionization-cooling.
Several studies are in progress and results are in preparation for publication.

\section{Acknowledgement}
Work described here has been made possible through generous funding from the Department of Energy and National Science 
Foundation (USA), the Istituto Nazionale di Fisica Nucleare (Italy), the Science and Technology Facilities Council (UK), the European Community under the European Commission Framework Programme~7, the Japan Society for the Promotion of Science and the Swiss National Science Foundation.


\vspace{4mm} 

\begin{thebibliography}{99}   % Use for  10-99  references
%\begin{thebibliography}{9} % Use for 1-9 references

  % Introduction
  \bibitem{geer} S.~Geer, ``Neutrino beams from muon storage rings: Characteristics and physics potential''. Phys.\ Rev.\ D {\bf 57} (1998) 6989
  \bibitem{neuffer} D.~Neuffer, ``Principles and applications of muon cooling'', Part.\ Accel.\ 14 (1983) 75
  \bibitem{stratakis} D.~Stratakis and R. Palmer, Phy.\ Rev.\ ST Accel. Beams 18 (2015) 031003
  \bibitem{fernow} R.~C.~Fernow and J.~C.~Gallardo. ``Muon transverse ionization cooling: Stochastic approach''. Phys.\ Rev.\ E {\bf 52}, 1039 (1995)
  \bibitem{blackmore} V.~Blackmore, ``Recent results from the study of emittance evolution in MICE'', presented at IPAC'18, paper TUPML067.
  
  % Scattering
  \bibitem{muscat1} MuScat Collaboration, W. J. Murray, ``Comparison of MuScat data with GEANT4'', Nucl. Phys. Proc. Suppl. 149 (2005) p. 99-103
  \bibitem{muscat2} D. Attwood et al., ``The scattering of muons in low Z materials'',  Nucl. Instrum. Meth. B251 (2006) 41
  \bibitem{geant4} GEANT4 Collaboration, S. Agostinelli et al., ``GEANT4: A Simulation toolkit'', Nucl. Instrum. Meth. A506 (2003) p. 250-303
  \bibitem{moliere} G.~Moliere. Theory of the scattering of fast charged particles. 2. Repeated and multiple scattering. Z. Naturforsch., A3:78–97, 1948.
  \bibitem{cc} T.~Carlisle, ``Step IV of the Muon Ionization Cooling Experiment (MICE) and the multiple scattering of muons''. PhD thesis, Oxford U., 2013
  \bibitem{nufact} R.~Bayes, ``Measurements of the Multiple Coulomb Scattering of Muons by MICE'', NuFact16 (2016)
  \bibitem{rossi} B.~Rossi, K.~Greisen, ``Cosmic-Ray Theory'', Rev. Mod. Phys. 13, 240, 1941       
  
  % Energy loss
  \bibitem{bethe} S.~Peter, ``Particle penetration and radiation effects''. Springer Series in Solid State Sciences, 151. Berlin Heidelberg: Springer-Verlag
  
\end{thebibliography}
\end{document}